\documentclass[letter,11pt]{article}
\pdfoutput=1 
\usepackage{jcappub} 
\usepackage[T1]{fontenc} 
\usepackage{subfig}
\usepackage{wrapfig}
\newcommand{\myparallel}{{\mkern3mu\vphantom{\perp}\vrule depth 0pt\mkern2mu\vrule depth 0pt\mkern3mu}}
\newcommand{\rpp}{r_{\perp},r_{\myparallel}}
\newcommand{\mpcoh}{\,h^{-1}\,{\rm Mpc}}
\newcommand{\rnc}{}
\newcommand{\correction}{}
\usepackage{amssymb}
\usepackage{astrobib_mnras2e}
\title{Mitigating the Impact of the DESI Fiber Assignment
on Galaxy Clustering}

\author[a]{Angela Burden}
\author[a]{Nikhil Padmanabhan}
\author[b]{Robert N. Cahn}
\author[b,c,d]{Martin J. White}
\author[e]{Lado Samushia}

\affiliation[a]{Dept. of Physics, Yale University, New Haven, CT 06511, USA}
\affiliation[b]{Physics Division, Lawrence Berkeley National Laboratory,1 Cyclotron Road, Berkeley, CA 94720, USA}
\affiliation[c]{Department of Physics, University of California, Berkeley, California}
\affiliation[d]{Department of Astronomy, University of California, Berkeley, California}
\affiliation[e]{Physics Department, Kansas State University,116 Cardwell Hall,1228 N. 17th St. Manhattan, KS 66506}
\emailAdd{angela.burden@yale.edu}

\abstract{We present a simple strategy to mitigate the impact of an incomplete 
spectroscopic redshift galaxy sample as a result of fiber assignment and survey tiling. 
The method has been designed for the Dark Energy Spectroscopic Instrument 
(DESI) galaxy survey but may have applications beyond this. 

We propose a modification to the usual correlation function that nulls the almost purely angular modes
affected by survey incompleteness due to fiber assignment. Predictions of this modified statistic 
can be calculated given a model of the two point correlation function. 
The new statistic can be computed with a slight modification to the data catalogues
input to the standard correlation function code and does not incur any additional computational time.

Finally we show that the spherically averaged 
baryon acoustic oscillation signal is not biased by the new statistic.  
}
\begin{document}
\maketitle
\flushbottom

\section{Introduction}
\label{sec:intro}

Accurate galaxy redshift measurements are obtained from fiber-fed galaxy spectroscopic surveys
by observing the positions of recognised features in each galaxy's spectrum.
There have been many such surveys in the past 16 years 
\cite{York_SDSS_2000, 2dFGRS, Stoughton02_SDSS1,DEEP2,WiggleZ,Driver11,Beutler11, Dawson13,Guzzo14}, 
and several large spectroscopic redshift surveys are planned for the near future
\cite{DESI, WEAVE, 4MOST}.
Spectra are collected via optical fibers fed {\rnc from the target location 
on the focal plane of the telescope to a spectrograph}. Plates are used to hold fibers in place; 
these are tiled over the footprint of the survey.
Each plate represents one set of observations; fibers are reconfigured 
between  observations by switching plates or mechanically moving the positions of the fibers.

 
In fiber-fed surveys an upper limit to the spectroscopic sample completeness 
in high density regions is {\rnc consequently} constrained by the physical size of a fiber and the number of fibers available. 
Additionally, overlap regions in the tiling geometry influence sample completeness and can alter 
the measured clustering of the galaxy distribution.

To mitigate these effects one may correct the galaxy density field
to account for objects that do not get assigned fibers. 
For example, the BOSS survey BAO analyses \cite{Anderson14_DR9, Anderson14_DR10}, 
up-weighted nearest (spectroscopically observed) neighbours to missed objects.
This method assumes that close pairs in angular separation are associated in redshift.

Alternatively one may directly apply corrections to the {\rnc two}-point statistics of the data. For example  
\cite{Hawkins03} measure the ratio of the angular correlation functions of parent target sample to
observed sample and apply this directly as a weight to the galaxy-galaxy pairs of the
correlation function. This corrects the reduction in amplitude of the small scale clustering due to 
fiber collisions. A different method is used in \cite{Guo12} where {\rnc the authors} divide the galaxy population
into non-collided galaxies that will always be observed, \textit{population-one} and 
collided galaxies (where the galaxies are too close to simultaneously be assigned fibers), \textit{population-two}.
In regions where the survey tiles overlap, the galaxies in population-two
have a greater chance of being observed than those in regions covered by only one tile. They use the statistics of the
observed population-two galaxies in tile overlap regions to correct the statistics of the population-two galaxies 
in areas with lower completeness. They combine the weighted correlation functions of population-one and observed population-two
to correct for the non-observed set.
 
The algorithm we propose in this paper is a modification to the usual correlation function rather than an attempt to recover the lost data.
It has been designed 
specifically for the DESI survey to correct for the effects of tiling and fiber assignment.

In section \ref{sec:DESI} we give an overview of the DESI survey, the fiber assignment algorithm and the mocks we use for our analysis.
In section \ref{sec:FAA} we propose a modification to the usual correlation function that reduces the effects of fiber assignment 
and show the modification can be predicted given a model of the two point correlation function. 
We present the results of the statistic computed with the modified mock data catalogues and fit the BAO peak.

\section{The DESI Survey}\label{sec:DESI}
We start with an overview of the DESI survey, and then discuss the details of the fiber-assignment algorithm. 

\subsection{An overview of DESI}

The Dark Energy Spectroscopic Instrument (DESI) has been
designed with the primary goal of measuring the influence of dark energy on the expansion rate of the Universe via 
baryon acoustic oscillations (BAO) 
and {\rnc determining} the growth of structure {\rnc from} redshift-space distortions (RSD)
\cite{DESI_I,DESI_II}. Additionally the observed large-scale structure map will provide a tool to
measure the sum of neutrino masses and place constraints on theories of
modified gravity and inflation \cite{Font-Ribera14}.

The ground based instrument will measure the spectroscopic redshifts of objects over {\rnc five years}
via ten fiber-fed {\rnc three-arm} spectrographs \cite{Smee13}.
It has the capability of measuring 5000 spectra concurrently  with fibers located at target positions by 
robotically-actuated fiber positioners. The instrument will be installed on the 4-m Mayall Telescope at {\rnc Kitt Peak, Arizona}. 

The DESI survey will cover 14,000 deg$^2$ over which it will measure
the spatial distribution of four classes of objects identified from pre-existing and ongoing imaging data \cite{Flaugher15,Wright10,Arjun16}. 
The target classes are luminous red galaxies (LRGs), to be observed up to $z=1.0$, 
emission line galaxies (ELGs), observed up to $z=1.7$, and quasars (QSO).
Quasars will be observed as tracers of the dark matter and the Lyman Alpha absorption features 
of a high redshift sample of quasars, spanning $2.1<z<3.5$, will be used to probe the intergalactic medium.
A further target class, the bright galaxies
will utilise the telescope time when the moon is above the horizon and 
these bright objects will be observed to $z \approx 0.4$ \cite{Wechsler15}.

The precision required to measure the BAO scale, the growth of structure from RSDs and the sum of neutrino masses 
from the two-point correlation function of observed galaxy pairs means that our measurement must be unbiased.
However, the spectroscopic observation method described below, distorts the true two-point statistics of the galaxies.

\subsection{Tiling and Fiber Assignment}

\begin{figure*}[h!]
    \centering
 \resizebox{0.5\columnwidth}{!}{\fbox{\includegraphics{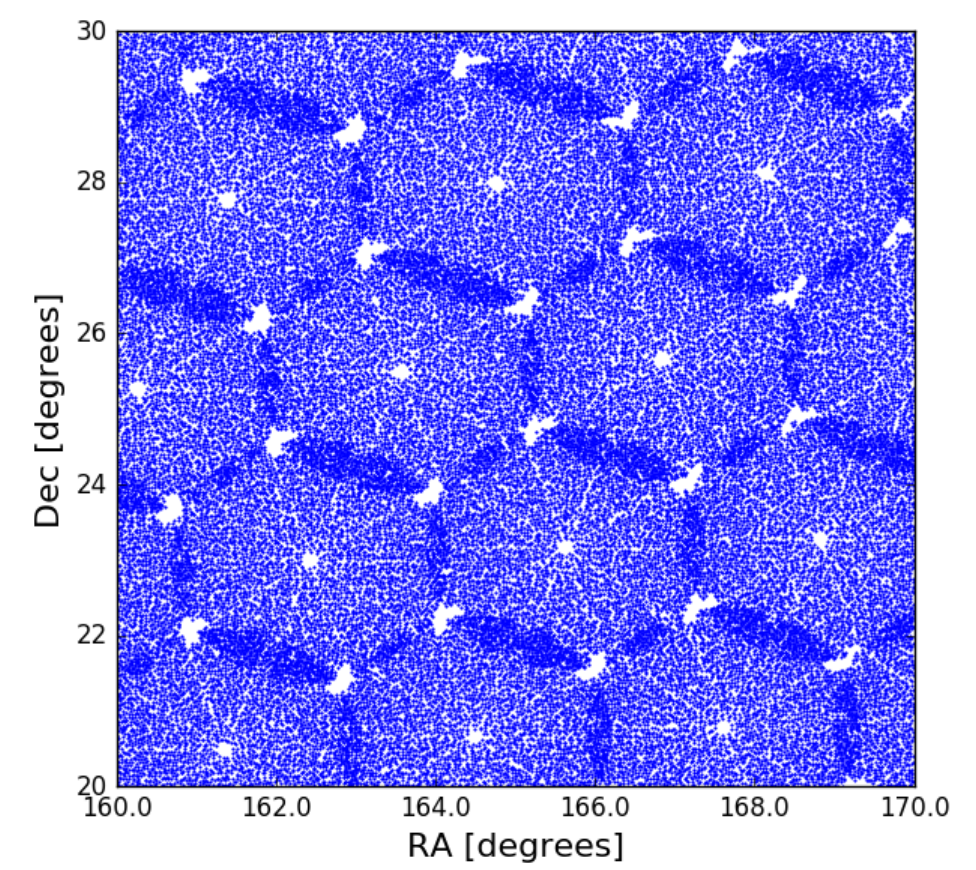}}}
  \resizebox{0.488\columnwidth}{!}{\fbox{\includegraphics{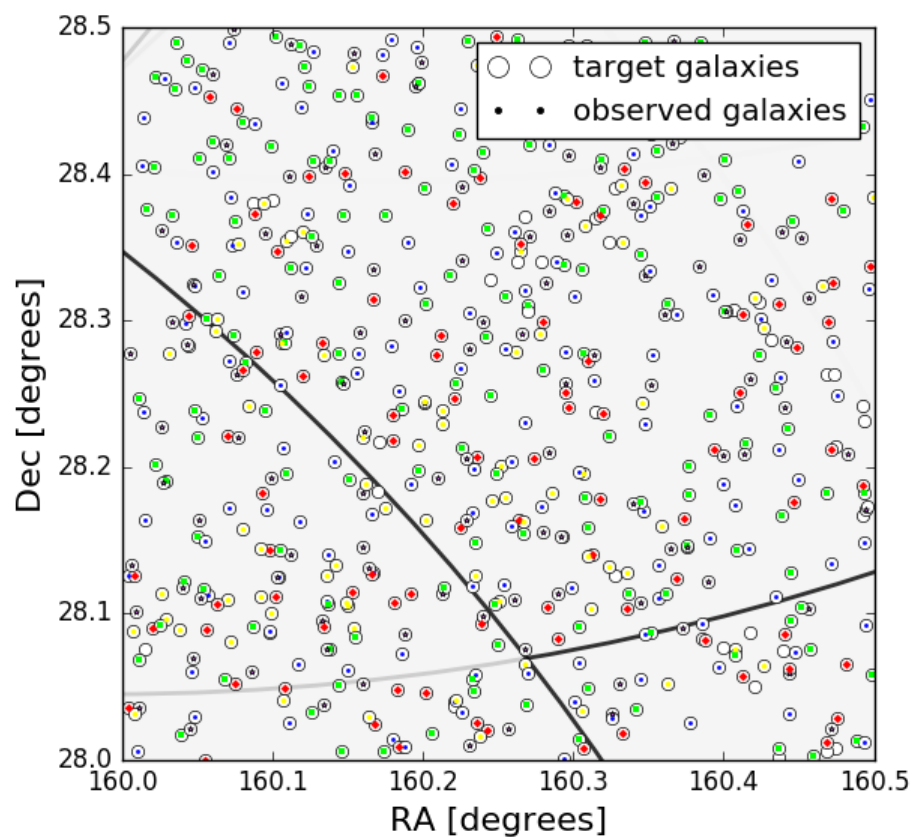}}}
    \caption{The left panel shows a ten by ten degree patch of the survey. The blue dots are the simulated ELG observations after pass-1. The tiling pattern
    is imprinted in the distribution. The right hand panel shows a 1/2 by 1/2 degree patch of the survey. The positions of target galaxies are displayed as black circles.
    The blue, green, purple, red and yellow points show the simulated observations after pass-1 to 5 respectively. After all {\rnc five} passes the empty circles are the unobserved galaxies.}
  \label{fig:foot}
\end{figure*}

Approximately 2000 pointings (tiles) will cover the survey the footprint (one pass) in a year.
The tiles are offset at each pass allowing a total of 10,000 unique tiled observations over five years.
Five thousand robotic positioners are located on the focal plane of the telescope which has an instrumented area of about 7.5 sq. deg.
Each positioner carries an optical fiber and on completion of an observation, fibers are mechanically moved to a new target position on the tile
ready for the next set of observations.

The robotically actuated fiber positioner will be programmed to {\rnc select} targets according 
to an algorithm {\rnc that} we refer to as the fiber assignment algorithm. Not all DESI targets will be assigned a fiber.
The algorithm chooses targets based on their priority, the fiber reach, 
 {\rnc the number of observations required and the need to avoid collisions between positioners. }
 
 The classes of objects to be observed in priority order during dark time are 
Lyman-alpha QSOs, QSOs, LRGs and ELGs. 
{\rnc Consequently} after {\rnc five} passes $\sim80\%$ completeness is expected for the lowest priority ELG sample 
compared to $\sim100\%$ for higher priority targets. 
\\
The clustering statistics of the ELG sample will be affected by
\begin{itemize}
\item Tile overlap; higher completeness in overlap regions.
\item Higher priority objects masking targets.
\item Fiber size and availability; high density regions get under-sampled due to 
physical limitations of the fibers.
\item Interference from QSO targets which may have correlated overdensities to the ELGs.
\end{itemize} 

The left panel of figure \ref{fig:foot} shows a 10 degree by 10 degree patch of the survey. The blue points are the 
simulated pass 1 observed ELG targets. The tiling pattern is clearly seen in the distribution. The right panel shows
a close up of target ELG galaxies within a {\rnc 1/2 degree by 1/2 degree patch} of the survey.
The target galaxies are shown as black circles and the different coloured symbols represent which of these galaxies 
are chosen to be observed by the fiber assignment algorithm at each pass. Empty circles show the missed galaxies.
The tiles are in grey outlined in black, although there are no full tiles shown in the figure.

In this work we have used a preliminary version of the DESI fiber assignment scheme \cite{Cahn15}.
The code takes as input
\begin{enumerate}
\item RA and Dec for all targets plus a priority {\rnc value},
\item RA and Dec for each tile centre,
\item the position of each of the 5000 fibers on the focal plane,
\item a synthetic spectroscopic redshift file to mimic the process of observation,
\item standard star and sky-fiber target files.
\end{enumerate}

The RA and Dec of all targets and each of the 10,000 tile centres are read in; 
each tile centre has an associated pass number (1-5).
The positions of the central body and fiber holder of the fiber positioners are 
computed to prevent collisions at the time of allocation.

Each fiber on each tile is passed a list of unique target IDs within reach of the fiber; targets are also 
designated a list of possible fibers.

The code assigns fibers sequentially for each pass by choosing objects with the highest 
observation priority. If multiple objects share the same priority, objects with the largest 
number of observations remaining are chosen or if the targets have equal priority and
number of observations left, one is chosen at random.
The target is marked as observed
and the number of remaining observations reflected in subsequent fiber assignment choices. 
Each tile is assigned 400 sky-fiber and 100 standard star fibers out of 5000. {\rnc These} will be used to {\rnc determine the sky brightness and calibrate 
the spectra}.

Once fibers have been assigned to targets, two redistribution processes take place. The first 
identifies targets within range of unused fibers, if a fiber currently assigned to one of those targets can be moved to a new target, this redistribution
takes place. This maximises the 
number of targets that get observed in total. The second process counts the fraction of unused fibers on each tile and adjusts the 
algorithm so that the fraction is approximately constant over the survey footprint.

Once the fibers have been redistributed, the synthetic real-time observations begin.
The synthetic file mimics the observation process and provides the spectroscopic information 
collected after each pass. The objects assigned 
to fibers are `observed' and assigned redshifts. 
If after observation the object is found to be something different from the target class, 
the observation plan for subsequent passes 
is updated.

\begin {table}
\caption {Percentage completeness after each pass} \label{tab:completeness} 
\begin{center}
\begin{tabular}{| l || c | c | c | c | c | }
    \hline
    pass no. & 1 & 2 & 3 & 4 & 5 \\ \hline 
    ELG completeness (\%) & 23 & 44 & 61 & 73 & 81  \\ \hline 
    \hline
\end{tabular}
\end{center}
\end{table}

Table \ref{tab:completeness} lists the completeness of the ELGs in 
one of the synthetic catalogues after each pass. The completeness of target samples increases 
with pass number and at {\rnc pass-5}, spectroscopic redshifts 
for $\sim80\%$ of the ELG sample are expected.

In the above observation strategy, the fiber assignment algorithm will have a greater impact 
on the clustering signal at lower pass numbers 
where the sample is less complete. 
{\rnc Studying the results after just the first pass accentuates the impact of fiber assignment on the two-point correlation function.}

For efficiency in the subsequent analysis a simplified version of the code is run on the mock catalogues. 
This bypasses the redistribution processes. However we find the same
conclusions as presented in this paper when running the full code on target files with all types
of objects.

\subsection{Mocks}\label{sec:mocks}

We use a set of 25 Quick Particle Mesh (QPM) \cite{White14} mock galaxy catalogues
to perform our analyses. 
They are designed specifically for DESI and mimic the survey footprint and 
radial selection function.
These mock catalogues are run though the fiber assignment 
pipeline; the resulting {\rnc two-point} statistics are shown in section \ref{sec:FAA}.

The QPM mocks are created by generating a low resolution particle mesh 
using an N-body code.
The initial conditions are set using second order Lagrangian Perturbation Theory (2LPT)
at $z$=25. At each step the force is computed using Fast Fourier Transforms (FFTs) and
a subset of particles (chosen based on local density) are selected as dark matter halos
and assigned a halo mass.
The mass values are tuned such that the mass function and large-scale bias 
matches that of higher resolution simulations.
Halos are populated with galaxies using the Halo Occupation Distribution (HOD) function
of \cite{Tinker12}. The ELG HOD is tuned such that the small scale projected correlation function matches 
that of the blue galaxies with a high star formation rate measured in the DEEP2 survey \cite{Mostek13}. 
The same procedure is carried out for the LRGs {\correction such that they have the same power law correlation function as in \cite{White11}.}
The model in \cite{Conroy13} is used 
to create the QSO population.

To compute our {\rnc two-point} statistics, each mock requires 
a corresponding random catalogue that traces the footprint of the mock but 
has no spatial correlation between points.
To construct the random catalogue, 
random RA and Dec positions are assigned to particles within the survey footprint.
The particles are assigned redshifts so that the random catalogue has the same radial 
selection function as the {\correction average mock selection function} at the $\Delta z =0.026 $ level.
We chose the $\Delta z$ bin to be large enough to contrast with natural fluctuations
in the mock galaxy density field along the line of sight. This random catalogue is
our parent random catalogue that we use to compute the {\rnc correlation function} of the sample 
pre-fiber assignment.
We run the random catalogue through
the fiber assignment algorithm to mimic the variations in completeness
in the galaxy sample due to tiling and the physical constraints of the fibers. We call this the
{\rnc fiber-assigned} random catalogue and use it to compute the {\rnc correlation function} of the {\rnc fiber-assigned} 
samples. 

\section{Fiber Assignment Artifacts}\label{sec:FAA}

To show the effects of the fiber assignment/tiling and test our method we run 25 QPM
 survey mock galaxies through the fiber assignment algorithm.
We use the publicly available CUTE correlation function code \cite{Alonso12} to compute 
{\rnc two-point} statistics in configuration space.
We chose to use the minimum-variance Landy-Szalay estimator \cite{Landy93}
\begin{equation}
\xi\left(r_{\perp}, r_{\myparallel}\right) =
\frac{DD\left(r_{\perp}, r_{\myparallel}\right) -2DR\left(r_{\perp}, r_{\myparallel}\right)
+RR\left(r_{\perp}, r_{\myparallel}\right)}{RR\left(r_{\perp}, r_{\myparallel}\right)},
\end{equation}
with 0.25$\mpcoh$ bin widths for $r_{\perp}$ and $r_{\myparallel}$ representing the separation distances
perpendicular and parallel to the line of sight between two objects.
We estimate the {\rnc one-dimensional} multipoles of the data with a Legendre polynomial compression
of the {\rnc two-dimensional} correlation function computed with separation $r, \mu$ where
\begin{equation}
r = ( r_{\perp}^2 + r_{\myparallel}^2)^{1/2} \phantom{xxxxxx} \mu = \frac{r_{\myparallel}}{r},
\end{equation}
so that 
\begin{equation}\label{eq:legendre}
\xi_{\ell}(r)\approx \frac{2\ell+1}{2}\sum_j\Delta\mu_j \xi(r,\mu_j)L_{\ell}(\mu_j),
\end{equation}
where $L_{\ell}$ is the Legendre polynomial of order $\ell$.

\begin{figure*}[h!]
    \resizebox{1.0\columnwidth}{!}{\includegraphics{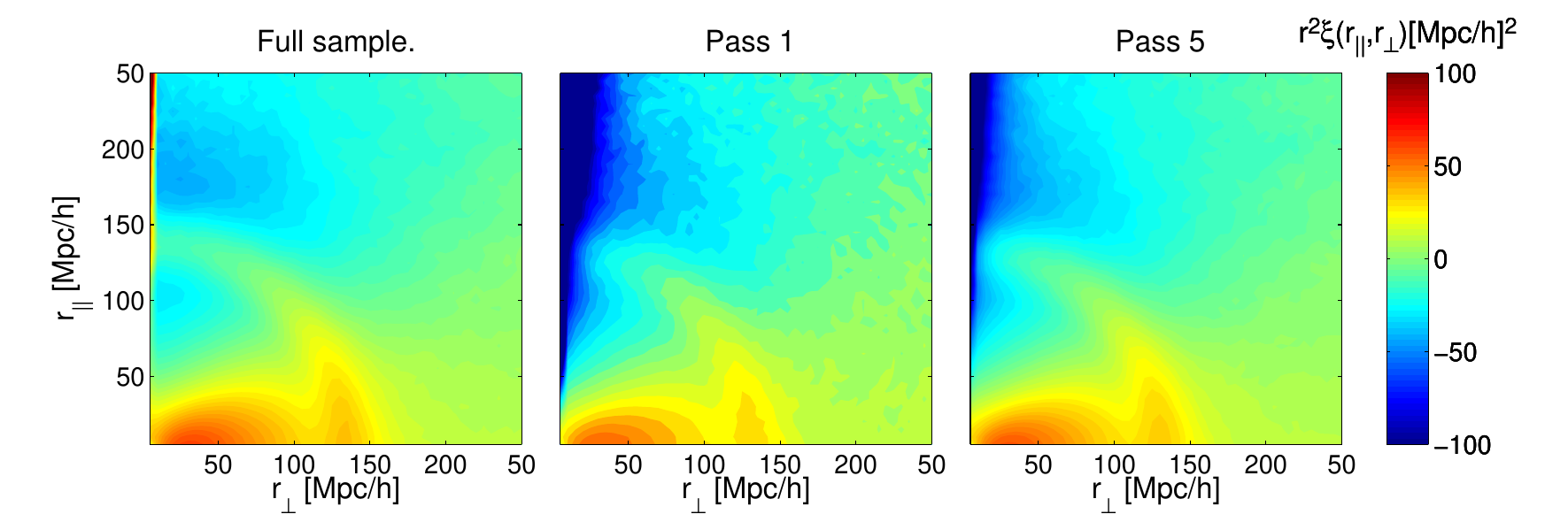}}
    \caption{Left, the two-dimensional correlation function of the complete sample, centre, the two-dimensional correlation function after 1 pass of the fiber-assignment algorithm computed with fiber-assigned randoms, right, the two-dimensional correlation function after 5 passes of the fiber-assignment algorithm computed with fiber-assigned randoms. The correlation functions of the fiber-assigned data look different to the full sample on scales of $r_{\perp}\lesssim 50 \mpcoh$ where pairs of galaxies with close angular proximity are not observed.}
  \label{fig:2Dps}
\end{figure*}

Figure \ref{fig:2Dps} shows the impact of the fiber-assignment and tiling averaged over the 25 mock data catalogues. Note we are using the 
{\rnc fiber-assigned} randoms as the default random catalogue to compare to {\rnc fiber-assigned} data. When constructing the fiber assigned random catalogue, we ran 4 random catalogues of the same density as the target ELG data through the algorithm and concatenated the output. {\correction We used 4 times the number of randoms than data point as this meant running 4 random catalogues (the same size as the mock data) through the fiber-assignment algorithm. To make cosmological measurements we suggest using at least 10 times the number of random points than data points but this was not necessary in our case. }
Tests using a random catalogue that is uniform in the RA, Dec plane within the 
footprint of the survey show much larger distortions when compared to the {\rnc fiber-assigned} data.
The plots show two dimensional correlation functions $\xi\left(r_{\perp}, r_{\myparallel}\right)$
for the average of the 25 mocks.
The plot on the left shows the pre-fiber assignment `true' sample {\rnc computed with all of the randoms}, the centre 
plot is the sample after 1 pass of the fiber-assignment algorithm the right is the sample after 5 passes of the fiber-assignment algorithm.

The plots show how the combined effect of the fiber assignment algorithm and tiling 
alter the line of sight distribution of pairs of galaxies with small separations ($r_{\perp}\lesssim 50 \mpcoh$)
perpendicular to line of sight compared to the original correlation function on the left.

\begin{figure*}[h!]
    \resizebox{1.0\columnwidth}{!}{\includegraphics{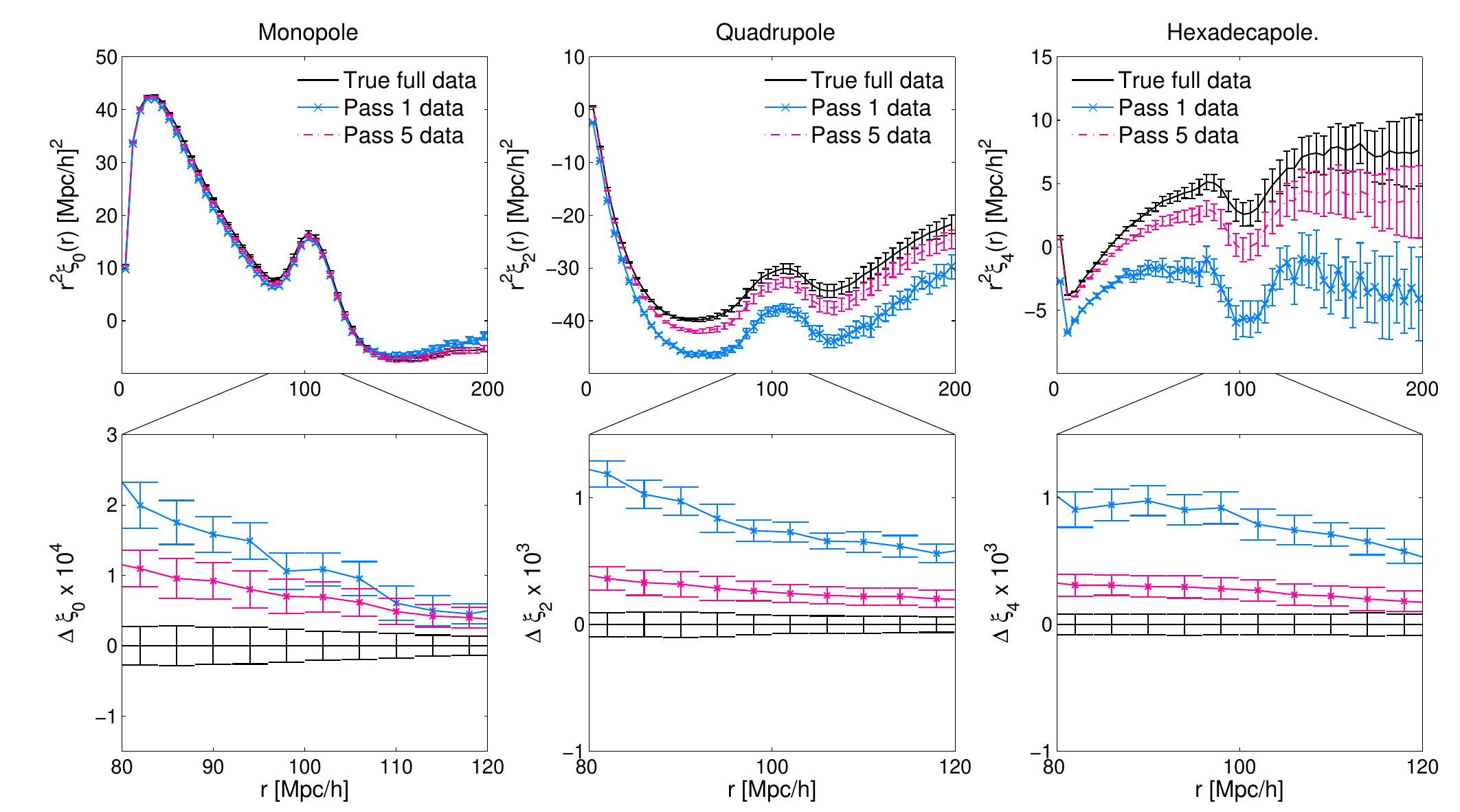}}
    \caption{The top panel of the plot shows the monopole (left), quadrupole (centre) and hexadecapole (right) of the complete sample (black), the pass-1 fiber-assignment data (blue cross) and fiber-assigned pass-5 data (pink dots). The bottom panel shows the difference between the pass-1 and original statistics (blue) and the pass-5 and original statistics (pink). {\correction The BAO region is isolated by showing the 80-120 $\mpcoh$ range}. From left to right the panels show the monopole, quadrupole and hexadecapole. The error bars show the standard deviation of the 25 mock catalogues.}
  \label{fig:diffmonoquad}
\end{figure*}

The monopole, quadrupole and hexadecapole of these samples are shown in the top panel of the top plot of figure \ref{fig:diffmonoquad}.
The monopole (left) of the {\rnc pass-5} data (pink dots) seems to reproduce the complete sample (black full line) but the {\rnc pass-1} data (blue crosses) diverge at small and large scales.
{\correction The lower panel of the top plot zooms in on the BAO scale (80-120 $\mpcoh$) and shows the differences between the complete sample statistics and the {\rnc pass-5} and {\rnc pass-1} data. 
A similar pattern emerges for the difference in the monopole, quadrupole and hexadecapole showing a change of shape of the function in the BAO region of the {\rnc fiber-assigned} data when compared to the full data set.}

\subsection{A Modified Algorithm}\label{sec:method}

To repair the structure in the tiled and {\rnc fiber-assigned} data a \textit{shuffled} random catalogue is created with exactly the same 
angular clustering as the data but randomly distributed along the line of sight retaining the 
radial selection function of the survey.
{\correction This shuffling method was previously used to produce redshifts for the BOSS random catalogue \cite{Ross12}.}
We replace the random catalogue in the numerator of the correlation function with the shuffled randoms, 
making our new estimator
\begin{equation}\label{eq:modified_data}
\tilde{\xi}\left(r_{\perp}, r_{\myparallel}\right) =
\frac{DD\left(r_{\perp}, r_{\myparallel}\right) -2DS\left(r_{\perp}, r_{\myparallel}\right)
+SS\left(r_{\perp}, r_{\myparallel}\right)}{RR\left(r_{\perp}, r_{\myparallel}\right)},
\end{equation}
where $S$ is the shuffled random catalogue.
 
\begin{figure*}[h!]
    \centering
    \resizebox{1.0\columnwidth}{!}{\includegraphics{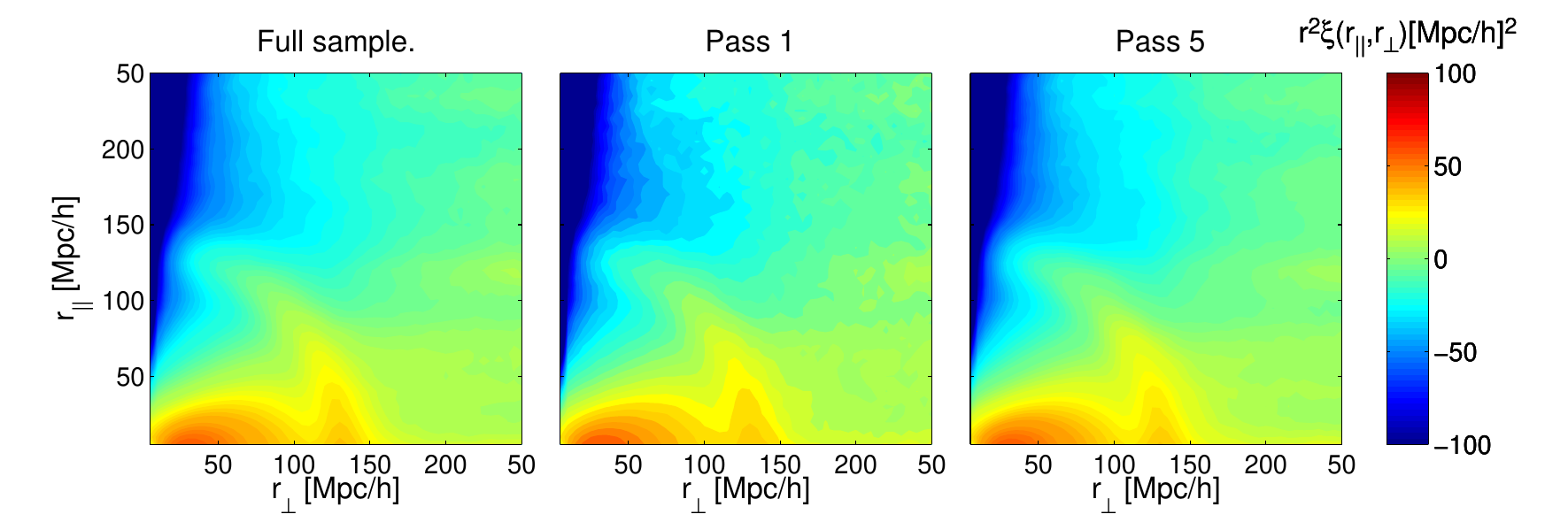}}
    \caption{Left, the modified {\rnc two-dimensional} correlation function of the complete sample, centre, the modified {\rnc two-dimensional} correlation function of post-fiber
    assignment {\rnc pass-1} data, right, the modified {\rnc two-dimensional} correlation function of the post-fiber assignment {\rnc pass-5} data. All three correlation functions use a shuffled random catalogue and show the same structure. }
  \label{fig:mod2Dps}
\end{figure*}

The {\rnc two-dimensional} correlation function of the new statistic is shown before (left) and after fiber assignment
({\rnc pass-1} centre, {\rnc pass-5} right) in figure \ref{fig:mod2Dps}. Note that the left hand plot of this figure is different from that in figure \ref{fig:2Dps} as the 
full sample statistic has been computed with the shuffled random catalogue.
The {\rnc two-dimensional} {\rnc two-point} correlation function of the {\rnc pass-1} and {\rnc pass-5} data computed with the shuffled random catalogue by eye
match those of the whole sample which has the modified feature at small $r_{\perp}$ along the line of sight.
To see more clearly the difference between the three data sets we compute the monopole, quadrupole and hexadecapole as before.

\begin{figure*}[h!]
    \resizebox{1.0\columnwidth}{!}{\includegraphics{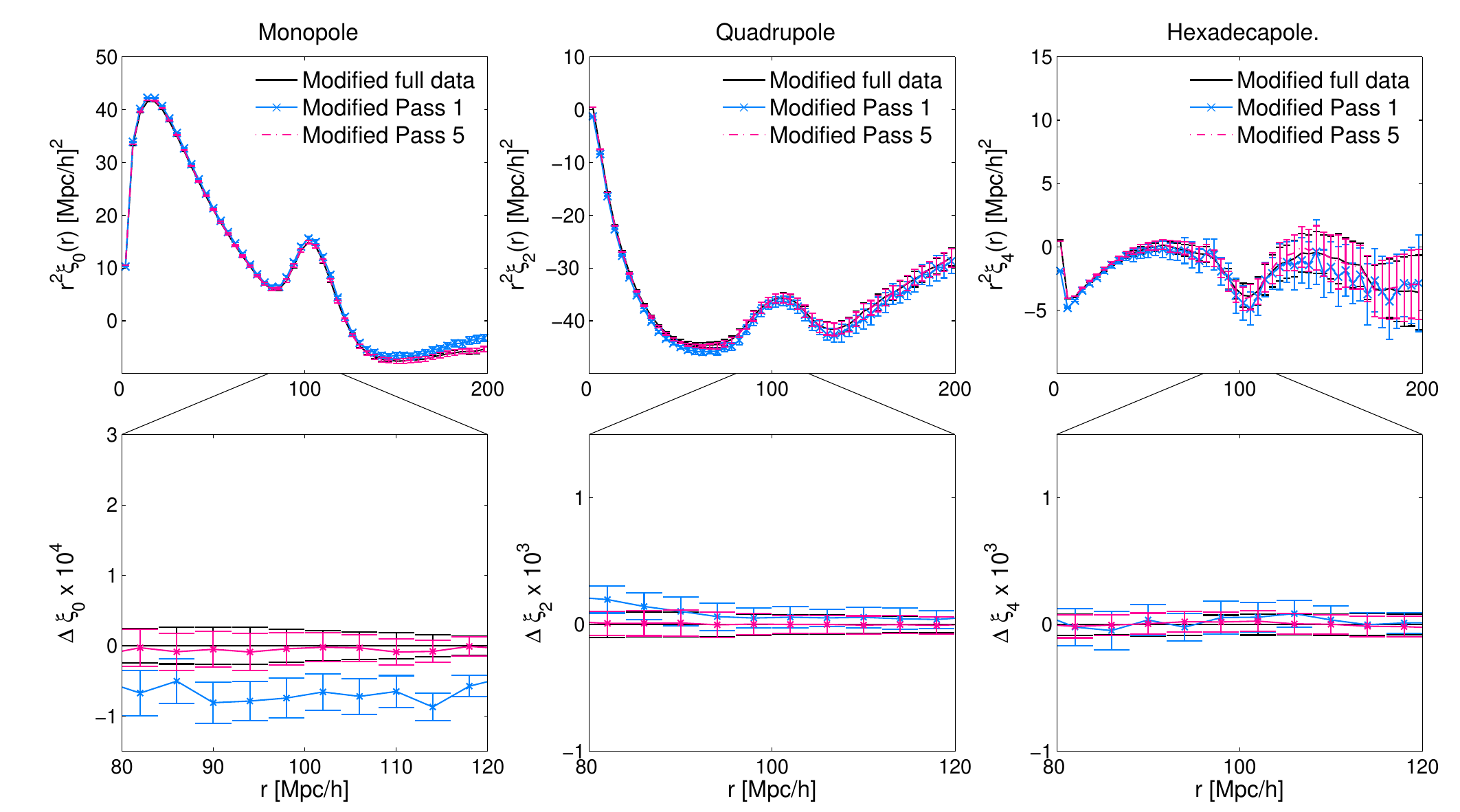}}
    \caption{The top panel of the plot shows the monopole (left) and quadrupole (centre) and hexadecapole (right) of the modified sample (black), the pass-1 fiber-assignment modified statistic (blue cross) and fiber-assigned pass-5 modified statistic (pink dots).The lower panel of the plot shows the difference between the pass-1 and original statistics (blue) and the pass-5 and original statistics (pink).{\correction The BAO region is isolated by showing the 80-120 $\mpcoh$ range}. From left to right the panels show the monopole, quadrupole and hexadecapole. The error bars show the standard deviation of the 25 mock catalogues. Comparing the bottom panel to that of figure \ref{fig:diffmonoquad}, {\correction it can be seen that the shape difference between the full sample and fiber-assigned samples over the BAO range is much smaller when using the modified correlation functions.}}
  \label{fig:mod_monoquad}
\end{figure*}

The upper panel of the plot in figure \ref{fig:mod_monoquad} shows the modified monopole, quadrupole and hexadecapole 
of the complete sample (black full line), {\rnc pass-1} data (blue crosses) and 
{\rnc pass-5} data (pink dots). The statistics of the modified sample at {\rnc pass-1} and {\rnc pass-5} seem to recover a noisier version of the complete sample suggesting that the
modified statistic approximately removes the effects of the tiling and fiber assignment.
The lower panels of the plot show the differences in the data sets. Although the differences show divergence from the original sample at small scales as with the
non-modified version, the differences are an order of magnitude smaller and the divergence occurs over a shorter separation range.

As the systematic effects we are removing are coupled to the angular components of the overdensity,
 to use the modified statistic we must understand the information being lost. Note that we do not compare the modified correlation function computed using the shuffled randoms to the complete sample (with no fiber assignment) as they are different statistics. 

\subsection{Understanding the Algorithm}

The modified correlation function consists of the galaxy density field ($n$), the shuffled random
catalogue ($\tilde{n}$) and the standard random catalogue ($\bar{n}$).
We can construct the modified density perturbation, $\tilde{\delta}$ as
\begin{equation}
 \tilde{\delta}({\bf r})\equiv \frac{n({\bf r}) - \tilde{n}({\bf r})}{\bar{n}({\bf r})},
 \end{equation}
 where the shuffled random catalogue contains the angular positions of the galaxies integrated over 
 the survey redshift range. The projected density is then redistributed along the line of sight in a ratio matching the ratio of 
 randoms as a function of redshift,
 \begin{equation}
 \tilde{n}({\bf r}) = \frac{\int n({\bf \gamma},\chi')d\chi' \int \bar{n}({\bf \gamma}',\chi )d{\bf \gamma'}}
 {\int \int  \bar{n}({\bf \gamma}', \chi')d{\bf \gamma}'d\chi' },
 \end{equation}
where $\gamma$ is the {\rnc two-dimensional} angular coordinate and $\chi$ the {\rnc line-of-sight} coordinate.
In terms of $\delta$ the modified overdensity can be written as
 \begin{equation}
 \tilde{\delta}({\bf r}) =\delta({\bf r}) - \frac{\int \delta({\bf \gamma},\chi')\bar{n}(\chi')d{\chi'}}{\int \bar{n}(\chi')d\chi'},
 \end{equation}
 where we have assumed that the random catalogue is constant in ${\bf \gamma}$. We note that this method is the configuration space equivalent to nulling the power in the $k_{\myparallel}=0$ bin in the power spectrum (Pinol et al. in prep).
 The modified correlation function is the ensemble average of the modified density fluctuations separated by distance $\mathbf{r - r'}$ 
 and can be expressed as
\begin{equation}\label{eq:sum}
\begin{split}
 \langle \tilde{\delta}({\bf r})  \tilde{\delta}({\bf r'})\rangle &=
  \left\langle\delta({\bf r})  \delta({\bf r'})\right\rangle  \\
 & - 2 \left\langle \delta({\bf r})\frac{\int \delta({\bf \gamma},\chi')\bar{n}(\chi')d\chi'}{\int \bar{n}(\chi')d\chi'}\right\rangle \\
&+ \left\langle \frac{\int \delta({\bf \gamma},\chi')\bar{n}(\chi')d\chi'}{\int \bar{n}(\chi')d\chi'} \frac{\int \delta({\bf \gamma}',\chi'')\bar{n}(\chi'')d\chi''}{\int \bar{n}(\chi'')d\chi''}\right\rangle.
\end{split}
\end{equation}
To compute the second term on the right hand side of the above equation we write
\begin{equation}
\left\langle \delta({\bf r})\frac{\int \delta({\bf \gamma},\chi')\bar{n}(\chi')d\chi'}{\int \bar{n}(\chi')d\chi'}\right\rangle = \frac{\int \xi(\theta,\chi,\chi') \bar{n}(\chi')d\chi'}{\int \bar{n}(\chi')d\chi'},
\end{equation}
where $\theta = | \gamma -\gamma' |$ and $\xi(\theta,\chi,\chi')$ is the correlation function between pairs of galaxies separated along the line-of-sight by distance $\chi -\chi'$.
We make the assumption that $\xi(\theta,\chi,\chi') \bar{n}(\chi')d\chi' \approx w(\theta)\bar{n}(\chi')d\chi'$, i.e. the angular correlation function, therefore the second term in equation \ref{eq:sum} is two times the angular correlation function.
Using the Limber approximation \cite{Limber53} we can write out the angular correlation function as
\begin{equation}
w(\theta) \approx \int \bar{n}^2(\chi') \int \xi(\rpp') dr'_{\myparallel} d\chi' \left(\int \bar{n}(\chi') d \chi' \right)^{-2}.
\end{equation}
Finally equation \ref{eq:sum} becomes
\begin{equation}\label{eq:sum2}
\begin{split}
 \langle \tilde{\delta}({\bf r})  \tilde{\delta}({\bf r'})\rangle &=
  \left\langle\delta({\bf r})  \delta({\bf r'})\right\rangle  \\
 & - 2 \int \bar{n}^2(\chi') \int \xi(\rpp') dr'_{\myparallel} d\chi' \left(\int \bar{n}(\chi') d \chi' \right)^{-2} \\
&+ \left\langle \frac{\int \delta({\bf \gamma},\chi')\bar{n}(\chi')d\chi'}{\int \bar{n}(\chi')d\chi'} \frac{\int \delta({\bf \gamma}',\chi'')\bar{n}(\chi'')d\chi''}{\int \bar{n}(\chi'')d\chi''}\right\rangle.
\end{split}
\end{equation}
The last term in the equation becomes negligible when the survey covers a large comoving distance
in which case the model of the corrected correlation function can be written as
\begin{equation}\label{eq:model}
  \tilde{\xi}(\rpp)=\xi(\rpp)  - 2\iint \xi(\rpp') \bar{n}^2(\chi)dr_{\myparallel}' d\chi \left(\int \bar{n}(\chi) d \chi \right)^{-2}.
\end{equation}
The correction becomes smaller as the radial range of the survey increases and as the radial range tends towards zero 
the full correction becomes equivalent to the original correlation function.

\subsection{Modelling the modified correlation function}\label{sec:model}
  
\begin{figure*}[h!]
    \centering
    \resizebox{1.0\columnwidth}{!}{\includegraphics{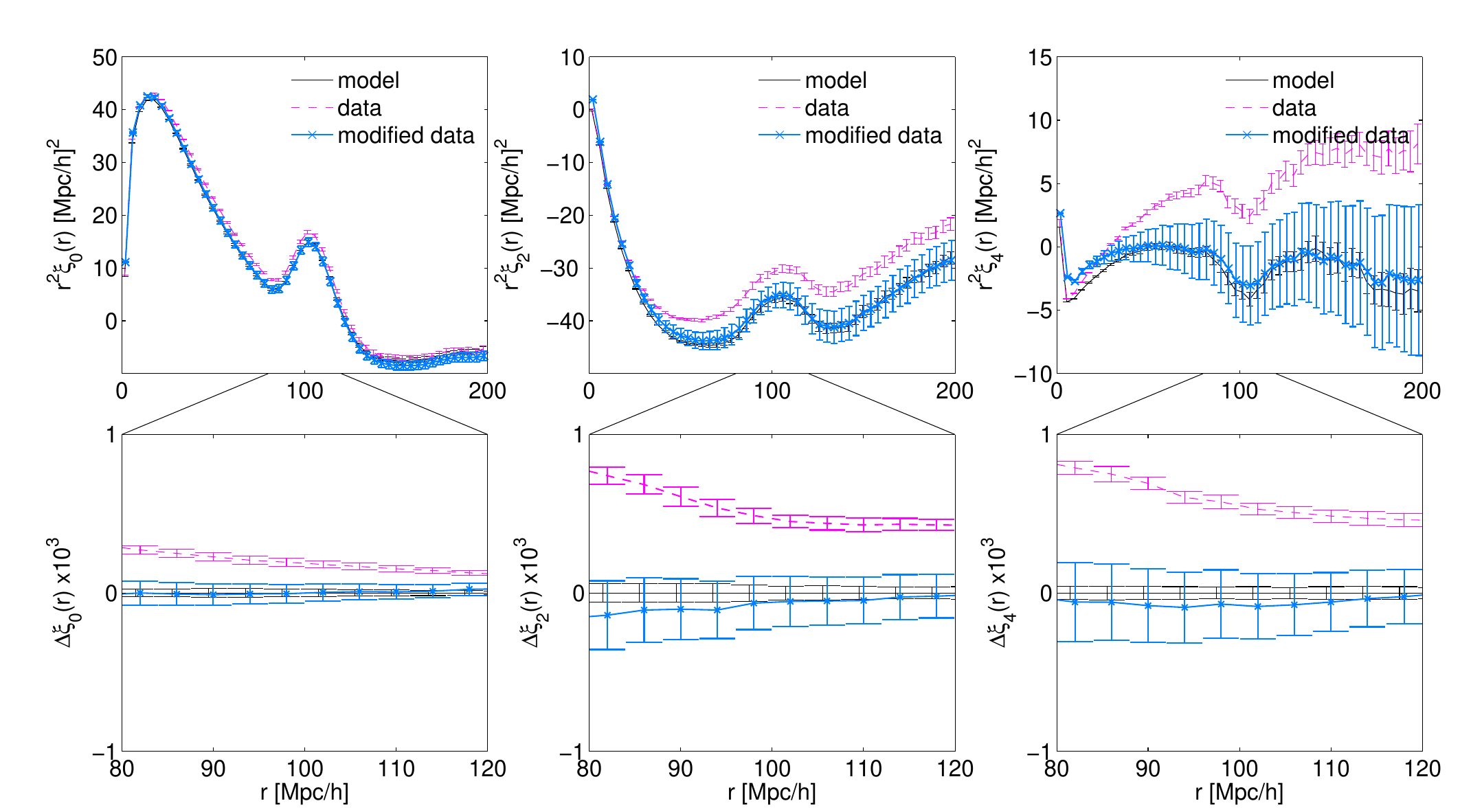}}
    \caption{{\correction The top panel of the plot shows the average monopole (left), quadrupole (centre) and hexadecapole (right) of the 25 mocks for the model in \ref{eq:model} (black line), (non-modified) full correlation function (pink dashed line) and the modified {\rnc two-point} statistics (shuffled randoms, blue crosses). The bottom panel shows the differences between them in the 80-120 $\mpcoh$ region. The error bars are the standard deviation of the mocks. The black line has error bars as the model is constructed from the correlation function of the full sample mocks. There are deviations between the model and the modified data on small scales however these are much smaller in the BAO region shown in the bottom panel. The non-modified correlation function however shows deviations from this line (in the bottom panel) that would not be captured by adjusting the broadband parameters as they change the shape of correlation functions in the BAO region shown.}}
  \label{fig:model}
\end{figure*}

In this subsection we compare our model (\ref{eq:model}) to the modified correlation function
measured in the {\correction full modified data set (with shuffled randoms)}.
Figure \ref{fig:model} shows the model, constructed as in equation \ref{eq:model}
using the $\xi(\rpp)$ of the full (non-{\rnc fiber-assigned}) sample.
The average of the model computed from the 25 mocks 
is shown as a black line and compared to the average 
modified {\rnc two-point} statistics (blue crosses) of the mock data computed with the
shuffled randoms.
The monopole and quadrupole by eye are
a very good match.
The hexadecapole shows a reasonable match but this inconsistencies may be due to the fact we use
$\xi \left(r_{\perp}, r_{\myparallel}\right)$ to construct the model and then translate to the 
$\mu, r$ coordinate system.

To get a quantitive comparison of the model and the data, we take the $\chi^2$ values 
\begin{equation}\label{eq:chi2}
\chi^2(\alpha) = [ {\bf d - m(\alpha)}]^{\textrm{T}} C^{-1}[ {\bf d - m(\alpha)}],
\end{equation}
where ${\bf d}$ is the modified correlation function vector of a mock as outlined in equation~\ref{eq:modified_data} and ${\bf m}$ is the model 
of the correlation function outlined in equation~\ref{eq:model} which has been computed from the full data set.
As we only have 25 mock catalogues we use the publicly available
covariance matrix of the DR11 CMASS BOSS galaxy sample \footnote{\url{https://www.sdss3.org/science/boss_publications.php}} and rescale it so that
\begin{equation}
\textrm{C}= \frac{V_{BOSS}}{V_{DESI}} \textrm{C}_{BOSS}
\end{equation}
where V$_{\textrm{BOSS}}$ = 10Gpc$^3$, and V$_{\textrm{DESI}}$ = 170 Gpc$^3$ are the volumes of
the BOSS (CMASS) and DESI survey respectively.

\begin{figure*}[h!]
\centering
    \resizebox{0.52\columnwidth}{!}{\includegraphics{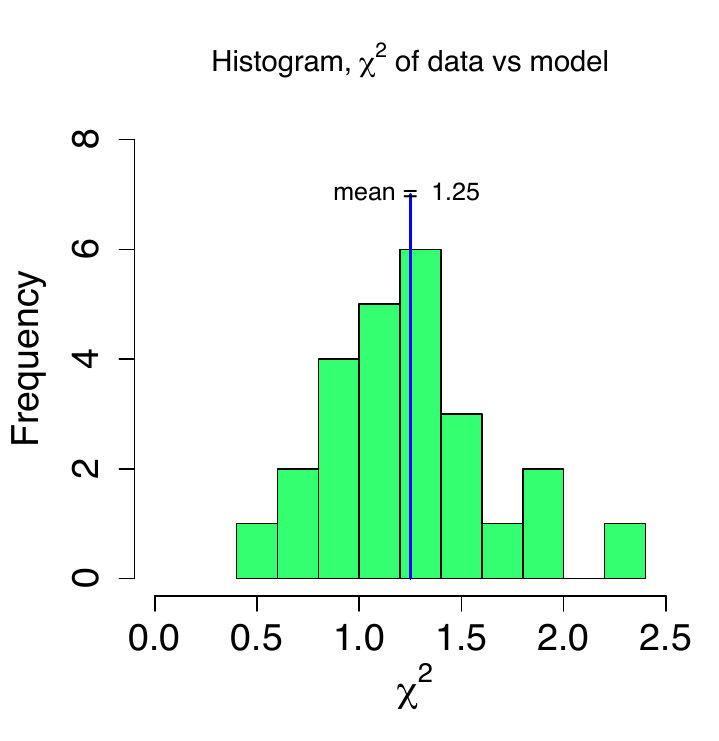}}
    \caption{The $\chi^2$ values computed for each mock when compared to the model (which is computed from the full data). The mean of the distribution is shown by the vertical line.    
    The $\chi^2$ values are higher than expected due to the conversion to the $r,\mu$ coordinate system where $\mu$ bins on small scales are coarsely sampled
    compared to the large scales.}
  \label{fig:chi2_model}
\end{figure*}

Figure \ref{fig:chi2_model} shows histogram $\chi^2$ values computed for each mock. The mean $\chi^2$ value is close to one. 
The statistic however also includes the errors introduced when converting the coordinate systems from $\rpp$ to $r,\mu$ and therefore we expect 
this to be a very conservative estimate.

To ascertain the usefulness of the model we follow \cite{Anderson14_DR9, Anderson14_DR10}
and fit the modified correlation functions of the 25 mock catalogues to the function

\begin{equation}\label{eq:xicor}
\textrm{B}^2\xi(r\alpha)_{\textrm{model}} + \frac{\textrm{a}_1}{r^2} + \frac{\textrm{a}_2}{r} + \textrm{a}_3,
\end{equation}
where a$_i$, are parameters that marginalise over the broadband shape and the B parameter adjusts 
the amplitude of the full correlation function.
The parameter that we are interested in is $\alpha$ as this encodes any shifts in the BAO feature compared to the original data. 
We increment $\alpha$ in bins of width $\Delta \alpha = 0.001$ in the range $0.95<\alpha<1.05$. The range we have chosen is small
as we know the true position of the BAO bump in our data and therefore we are not trying to measure cosmology but rather 
look for deviations from $\alpha=1$ and the variance of the modified correlation functions compared to the model.

For each $\alpha$ value we compute the
best fit values of the 4 parameters and compute the $\chi^2$ as in equation~\ref{eq:chi2} but where 
${\bf d}$ is the modified correlation function vector of a mock as before and ${\bf m}$ is the model 
of the correlation function with best fit parameters as in equation \ref{eq:xicor} for a given $\alpha$ value.


The mean $\alpha$ values computed from the modified correlation functions is
1.000, and the standard deviation of the mocks compared to the average best fit model is $\sigma _{\alpha}=0.0066$. The mean $\alpha$ from
the true correlation function is 1.000 (as expected) and the standard deviation is $\sigma _{\alpha}=0.0067$.
We therefore conclude that the modified correlation function is unbiased with respect to recovery of $\alpha$ and the error on this value.

\section{Discussion}

In this paper we have described an approximation of
the tiling and fiber assignment algorithm that will be used in the DESI survey 
to place fibers on target objects and collect spectroscopic redshifts.

We have shown that the angular clustering coupled with the spectroscopic observation 
method alters the {\rnc two-point} statistics in early stages of the 
survey when sample completeness is low.
A simple method of mitigating these effects has been presented and tested on mock catalogues.
The method uses a modified correlation function that removes small scale angular clustering. The modified correlation
functions of the lower completeness data sets that mimic the year 1 and year 5 DESI ELG sample recover the modified statistics 
of the full sample with an order of magnitude improvement over the non-modified statistic.

We have demonstrated that although
information about the small scale angular clustering is lost, the modified correlation function,
monopole, quadrupole can be modelled. The lost information is not detrimental to our 
spherically averaged BAO cosmological analysis although we have not tested the 
effect of the model on the anisotropic BAO analysis as yet.
{\correction The study of RSD measurements and the analysis of the anisotropic BAO measurements are deferred to future work. 
The modified correlation function is only removing information perpendicular to the line of sight where $\mu =0$, thus the RSD signal should not be modified. 
However the RSD signal is related to the density of the local environment and high density peaks may be flattened 
using this method.}

{\correction By nulling the $k_{\myparallel}= 0$ modes, it is possible to mitigate systematic errors associated with imaging data that are restricted to this plane. Again this is worth further investigation.}

{\correction We have not investigated how the modified catalogues will affect the reconstruction technique \cite{Eisenstein07, Padmanabhan12, Burden14}. The modified catalogues alter the over-density in an
anisotropic way thus will distort the reconstruction displacement vectors. The effects should be investigated before applying reconstruction to the modified catalogues.} 

This new statistic will allow one to make robust cosmological measurements of the BAO feature from early
DESI data if the current observation plan is implemented. 

Although the method has been designed specifically for the DESI survey we note that it may be
useful in other fiber-fed spectroscopic surveys where the sample completeness
is limited in angular resolution due to hardware constraints or the observed sample clustering is 
affected by the tiling pattern.

\acknowledgments
We thank Uros Seljak, Lucas Pinol, Nick Hand, Julien Guy and Kyle Dawson for useful discussions.
AB and NP are supported in part by DOE DE-SC0008080.

This research is supported by the Director, Office of Science, Office of High Energy Physics of the U.S. Department of Energy under Contract No. 
DEÐAC02Ð05CH1123, and by the National Energy Research Scientific Computing Center, a DOE Office of Science User Facility under the same 
contract; additional support for DESI is provided by the U.S. National Science Foundation, Division of Astronomical Sciences under Contract No. 
AST-0950945 to the National Optical Astronomy Observatory; the Science and Technologies Facilities Council of the United Kingdom; the Gordon 
and Betty Moore Foundation; the Heising-Simons Foundation; the National Council of Science and Technology of Mexico, and by the DESI 
Member Institutions.  The authors are honored to be permitted to conduct astronomical research on Iolkam DuÕag (Kitt Peak), a mountain with 
particular significance to the Tohono OÕodham Nation. 

\bibliography{mitigate_v1}

\providecommand{\href}[2]{#2}\begingroup\raggedright\begin{thebibliography}{10}

\bibitem{York_SDSS_2000}
D.~G. {York}, J.~{Adelman}, J.~E. {Anderson}, Jr., S.~F. {Anderson},
  J.~{Annis}, N.~A. {Bahcall} et~al., \emph{{The Sloan Digital Sky Survey:
  Technical Summary}}, \href{http://dx.doi.org/10.1086/301513}{\emph{\aj} {\bf
  120} (Sept., 2000) 1579--1587},
  [\href{http://arxiv.org/abs/astro-ph/0006396}{{\tt astro-ph/0006396}}].

\bibitem{2dFGRS}
M.~{Colless}, G.~{Dalton}, S.~{Maddox}, W.~{Sutherland}, P.~{Norberg},
  S.~{Cole} et~al., \emph{{The 2dF Galaxy Redshift Survey: spectra and
  redshifts}},
  \href{http://dx.doi.org/10.1046/j.1365-8711.2001.04902.x}{\emph{\mnras} {\bf
  328} (Dec., 2001) 1039--1063},
  [\href{http://arxiv.org/abs/astro-ph/0106498}{{\tt astro-ph/0106498}}].

\bibitem{Stoughton02_SDSS1}
C.~{Stoughton}, R.~H. {Lupton}, M.~{Bernardi}, M.~R. {Blanton}, S.~{Burles},
  F.~J. {Castander} et~al., \emph{{Sloan Digital Sky Survey: Early Data
  Release}}, \href{http://dx.doi.org/10.1086/324741}{\emph{\aj} {\bf 123}
  (Jan., 2002) 485--548}.

\bibitem{DEEP2}
M.~{Davis}, S.~M. {Faber}, J.~{Newman}, A.~C. {Phillips}, R.~S. {Ellis}, C.~C.
  {Steidel} et~al., \emph{{Science Objectives and Early Results of the DEEP2
  Redshift Survey}},  in \emph{Discoveries and Research Prospects from 6- to
  10-Meter-Class Telescopes II} (P.~{Guhathakurta}, ed.), vol.~4834 of
  \emph{\procspie}, pp.~161--172, Feb., 2003.
\newblock \href{http://arxiv.org/abs/astro-ph/0209419}{{\tt astro-ph/0209419}}.
\newblock \href{http://dx.doi.org/10.1117/12.457897}{DOI}.

\bibitem{WiggleZ}
M.~J. {Drinkwater}, R.~J. {Jurek}, C.~{Blake}, D.~{Woods}, K.~A. {Pimbblet},
  K.~{Glazebrook} et~al., \emph{{The WiggleZ Dark Energy Survey: survey design
  and first data release}},
  \href{http://dx.doi.org/10.1111/j.1365-2966.2009.15754.x}{\emph{\mnras} {\bf
  401} (Jan., 2010) 1429--1452}, [\href{http://arxiv.org/abs/0911.4246}{{\tt
  0911.4246}}].

\bibitem{Driver11}
S.~P. {Driver}, D.~T. {Hill}, L.~S. {Kelvin}, A.~S.~G. {Robotham}, J.~{Liske},
  P.~{Norberg} et~al., \emph{{Galaxy and Mass Assembly (GAMA): survey
  diagnostics and core data release}},
  \href{http://dx.doi.org/10.1111/j.1365-2966.2010.18188.x}{\emph{\mnras} {\bf
  413} (May, 2011) 971--995}, [\href{http://arxiv.org/abs/1009.0614}{{\tt
  1009.0614}}].

\bibitem{Beutler11}
F.~{Beutler}, C.~{Blake}, M.~{Colless}, D.~H. {Jones}, L.~{Staveley-Smith},
  L.~{Campbell} et~al., \emph{{The 6dF Galaxy Survey: baryon acoustic
  oscillations and the local Hubble constant}},
  \href{http://dx.doi.org/10.1111/j.1365-2966.2011.19250.x}{\emph{MNRAS} {\bf
  416} (Oct., 2011) 3017--3032}, [\href{http://arxiv.org/abs/1106.3366}{{\tt
  1106.3366}}].

\bibitem{Dawson13}
K.~S. {Dawson}, D.~J. {Schlegel}, C.~P. {Ahn}, S.~F. {Anderson},
  {\'E}.~{Aubourg}, S.~{Bailey} et~al., \emph{{The Baryon Oscillation
  Spectroscopic Survey of SDSS-III}},
  \href{http://dx.doi.org/10.1088/0004-6256/145/1/10}{\emph{The Astrophysical
  Journal} {\bf 145} (Jan., 2013) 10},
  [\href{http://arxiv.org/abs/1208.0022}{{\tt 1208.0022}}].

\bibitem{Guzzo14}
L.~{Guzzo}, M.~{Scodeggio}, B.~{Garilli}, B.~R. {Granett}, A.~{Fritz},
  U.~{Abbas} et~al., \emph{{The VIMOS Public Extragalactic Redshift Survey
  (VIPERS). An unprecedented view of galaxies and large-scale structure at 0.5
  $<$ z $<$ 1.2}},
  \href{http://dx.doi.org/10.1051/0004-6361/201321489}{\emph{\aap} {\bf 566}
  (June, 2014) A108}, [\href{http://arxiv.org/abs/1303.2623}{{\tt 1303.2623}}].

\bibitem{DESI}
M.~{Levi}, C.~{Bebek}, T.~{Beers}, R.~{Blum}, R.~{Cahn}, D.~{Eisenstein}
  et~al., \emph{{The DESI Experiment, a whitepaper for Snowmass 2013}},
  {\emph{ArXiv e-prints} (Aug., 2013) },
  [\href{http://arxiv.org/abs/1308.0847}{{\tt 1308.0847}}].

\bibitem{WEAVE}
G.~{Dalton}, S.~C. {Trager}, D.~C. {Abrams}, D.~{Carter}, P.~{Bonifacio},
  J.~A.~L. {Aguerri} et~al., \emph{{WEAVE: the next generation wide-field
  spectroscopy facility for the William Herschel Telescope}},  in \emph{Society
  of Photo-Optical Instrumentation Engineers (SPIE) Conference Series},
  vol.~8446 of \emph{Society of Photo-Optical Instrumentation Engineers (SPIE)
  Conference Series}, Sept., 2012.
\newblock \href{http://dx.doi.org/10.1117/12.925950}{DOI}.

\bibitem{4MOST}
R.~S. {de Jong}, O.~{Bellido-Tirado}, C.~{Chiappini}, {\'E}.~{Depagne},
  R.~{Haynes}, D.~{Johl} et~al., \emph{{4MOST: 4-metre multi-object
  spectroscopic telescope}},  in \emph{Society of Photo-Optical Instrumentation
  Engineers (SPIE) Conference Series}, vol.~8446 of \emph{Society of
  Photo-Optical Instrumentation Engineers (SPIE) Conference Series}, Sept.,
  2012.
\newblock \href{http://arxiv.org/abs/1206.6885}{{\tt 1206.6885}}.
\newblock \href{http://dx.doi.org/10.1117/12.926239}{DOI}.

\bibitem{Anderson14_DR9}
L.~{Anderson}, E.~{Aubourg}, S.~{Bailey}, F.~{Beutler}, A.~S. {Bolton},
  J.~{Brinkmann} et~al., \emph{{The clustering of galaxies in the SDSS-III
  Baryon Oscillation Spectroscopic Survey: measuring D$_{A}$ and H at z = 0.57
  from the baryon acoustic peak in the Data Release 9 spectroscopic Galaxy
  sample}}, \href{http://dx.doi.org/10.1093/mnras/stt2206}{\emph{MNRAS} {\bf
  439} (Mar., 2014) 83--101}, [\href{http://arxiv.org/abs/1303.4666}{{\tt
  1303.4666}}].

\bibitem{Anderson14_DR10}
L.~{Anderson}, {\'E}.~{Aubourg}, S.~{Bailey}, F.~{Beutler}, V.~{Bhardwaj},
  M.~{Blanton} et~al., \emph{{The clustering of galaxies in the SDSS-III Baryon
  Oscillation Spectroscopic Survey: baryon acoustic oscillations in the Data
  Releases 10 and 11 Galaxy samples}},
  \href{http://dx.doi.org/10.1093/mnras/stu523}{\emph{MNRAS} {\bf 441} (June,
  2014) 24--62}, [\href{http://arxiv.org/abs/1312.4877}{{\tt 1312.4877}}].

\bibitem{Hawkins03}
E.~{Hawkins}, S.~{Maddox}, S.~{Cole}, O.~{Lahav}, D.~S. {Madgwick},
  P.~{Norberg} et~al., \emph{{The 2dF Galaxy Redshift Survey: correlation
  functions, peculiar velocities and the matter density of the Universe}},
  \href{http://dx.doi.org/10.1046/j.1365-2966.2003.07063.x}{\emph{\mnras} {\bf
  346} (Nov., 2003) 78--96}, [\href{http://arxiv.org/abs/astro-ph/0212375}{{\tt
  astro-ph/0212375}}].

\bibitem{Guo12}
H.~{Guo}, I.~{Zehavi} and Z.~{Zheng}, \emph{{A New Method to Correct for Fiber
  Collisions in Galaxy Two-point Statistics}},
  \href{http://dx.doi.org/10.1088/0004-637X/756/2/127}{\emph{\apj} {\bf 756}
  (Sept., 2012) 127}, [\href{http://arxiv.org/abs/1111.6598}{{\tt 1111.6598}}].

\bibitem{DESI_I}
{DESI Collaboration}, A.~{Aghamousa}, J.~{Aguilar}, S.~{Ahlen}, S.~{Alam},
  L.~E. {Allen} et~al., \emph{{The DESI Experiment Part I: Science,Targeting,
  and Survey Design}}, {\emph{ArXiv e-prints} (Oct., 2016) },
  [\href{http://arxiv.org/abs/1611.00036}{{\tt 1611.00036}}].

\bibitem{DESI_II}
{DESI Collaboration}, A.~{Aghamousa}, J.~{Aguilar}, S.~{Ahlen}, S.~{Alam},
  L.~E. {Allen} et~al., \emph{{The DESI Experiment Part II: Instrument
  Design}}, {\emph{ArXiv e-prints} (Oct., 2016) },
  [\href{http://arxiv.org/abs/1611.00037}{{\tt 1611.00037}}].

\bibitem{Font-Ribera14}
A.~{Font-Ribera}, P.~{McDonald}, N.~{Mostek}, B.~A. {Reid}, H.-J. {Seo} and
  A.~{Slosar}, \emph{{DESI and other Dark Energy experiments in the era of
  neutrino mass measurements}},
  \href{http://dx.doi.org/10.1088/1475-7516/2014/05/023}{\emph{\jcap} {\bf 5}
  (May, 2014) 023}, [\href{http://arxiv.org/abs/1308.4164}{{\tt 1308.4164}}].

\bibitem{Smee13}
S.~A. {Smee}, J.~E. {Gunn}, A.~{Uomoto}, N.~{Roe}, D.~{Schlegel}, C.~M.
  {Rockosi} et~al., \emph{{The Multi-object, Fiber-fed Spectrographs for the
  Sloan Digital Sky Survey and the Baryon Oscillation Spectroscopic Survey}},
  \href{http://dx.doi.org/10.1088/0004-6256/146/2/32}{\emph{The Astrophysical
  Journal} {\bf 146} (Aug., 2013) 32},
  [\href{http://arxiv.org/abs/1208.2233}{{\tt 1208.2233}}].

\bibitem{Flaugher15}
B.~{Flaugher}, H.~T. {Diehl}, K.~{Honscheid}, T.~M.~C. {Abbott}, O.~{Alvarez},
  R.~{Angstadt} et~al., \emph{{The Dark Energy Camera}},
  \href{http://dx.doi.org/10.1088/0004-6256/150/5/150}{\emph{\aj} {\bf 150}
  (Nov., 2015) 150}, [\href{http://arxiv.org/abs/1504.02900}{{\tt
  1504.02900}}].

\bibitem{Wright10}
E.~L. {Wright}, P.~R.~M. {Eisenhardt}, A.~K. {Mainzer}, M.~E. {Ressler}, R.~M.
  {Cutri}, T.~{Jarrett} et~al., \emph{{The Wide-field Infrared Survey Explorer
  (WISE): Mission Description and Initial On-orbit Performance}},
  \href{http://dx.doi.org/10.1088/0004-6256/140/6/1868}{\emph{\aj} {\bf 140}
  (Dec., 2010) 1868--1881}, [\href{http://arxiv.org/abs/1008.0031}{{\tt
  1008.0031}}].

\bibitem{Arjun16}
A.~Dey, D.~Rabinowitz, A.~Karcher, C.~Bebek, C.~Baltay, D.~Sprayberry et~al.,
  \emph{Mosaic3: a red-sensitive upgrade for the prime focus camera at the
  mayall 4m telescope},  2016.
\newblock 10.1117/12.2231488.

\bibitem{Wechsler15}
R.~H. {Wechsler} and {DESI Collaboration}, \emph{{The Dark Energy Spectroscopic
  Instrument (DESI): Bright-Time Science Program}},  in \emph{American
  Astronomical Society Meeting Abstracts}, vol.~225 of \emph{American
  Astronomical Society Meeting Abstracts}, p.~336.06, Jan., 2015.

\bibitem{Cahn15}
R.~N. {Cahn}, S.~J. {Bailey}, K.~S. {Dawson}, J.~{Forero Romero}, D.~J.
  {Schlegel}, M.~{White} et~al., \emph{{The Dark Energy Spectroscopic
  Instrument (DESI): Tiling and Fiber Assignment}},  in \emph{American
  Astronomical Society Meeting Abstracts}, vol.~225 of \emph{American
  Astronomical Society Meeting Abstracts}, p.~336.10, Jan., 2015.

\bibitem{White14}
M.~{White}, J.~L. {Tinker} and C.~K. {McBride}, \emph{{Mock galaxy catalogues
  using the quick particle mesh method}},
  \href{http://dx.doi.org/10.1093/mnras/stt2071}{\emph{\mnras} {\bf 437} (Jan.,
  2014) 2594--2606}, [\href{http://arxiv.org/abs/1309.5532}{{\tt 1309.5532}}].

\bibitem{Tinker12}
J.~L. {Tinker}, E.~S. {Sheldon}, R.~H. {Wechsler}, M.~R. {Becker}, E.~{Rozo},
  Y.~{Zu} et~al., \emph{{Cosmological Constraints from Galaxy Clustering and
  the Mass-to-number Ratio of Galaxy Clusters}},
  \href{http://dx.doi.org/10.1088/0004-637X/745/1/16}{\emph{\apj} {\bf 745}
  (Jan., 2012) 16}, [\href{http://arxiv.org/abs/1104.1635}{{\tt 1104.1635}}].

\bibitem{Mostek13}
N.~{Mostek}, A.~L. {Coil}, M.~{Cooper}, M.~{Davis}, J.~A. {Newman} and B.~J.
  {Weiner}, \emph{{The DEEP2 Galaxy Redshift Survey: Clustering Dependence on
  Galaxy Stellar Mass and Star Formation Rate at z \~{} 1}},
  \href{http://dx.doi.org/10.1088/0004-637X/767/1/89}{\emph{\apj} {\bf 767}
  (Apr., 2013) 89}, [\href{http://arxiv.org/abs/1210.6694}{{\tt 1210.6694}}].

\bibitem{White11}
M.~{White}, M.~{Blanton}, A.~{Bolton}, D.~{Schlegel}, J.~{Tinker}, A.~{Berlind}
  et~al., \emph{{The Clustering of Massive Galaxies at z \~{} 0.5 from the
  First Semester of BOSS Data}},
  \href{http://dx.doi.org/10.1088/0004-637X/728/2/126}{\emph{\apj} {\bf 728}
  (Feb., 2011) 126}, [\href{http://arxiv.org/abs/1010.4915}{{\tt 1010.4915}}].

\bibitem{Conroy13}
C.~{Conroy} and M.~{White}, \emph{{A Simple Model for Quasar Demographics}},
  \href{http://dx.doi.org/10.1088/0004-637X/762/2/70}{\emph{\apj} {\bf 762}
  (Jan., 2013) 70}, [\href{http://arxiv.org/abs/1208.3198}{{\tt 1208.3198}}].

\bibitem{Alonso12}
D.~{Alonso}, \emph{{CUTE solutions for two-point correlation functions from
  large cosmological datasets}}, {\emph{ArXiv e-prints} (Oct., 2012) },
  [\href{http://arxiv.org/abs/1210.1833}{{\tt 1210.1833}}].

\bibitem{Landy93}
S.~D. {Landy} and A.~S. {Szalay}, \emph{{Bias and variance of angular
  correlation functions}}, \href{http://dx.doi.org/10.1086/172900}{\emph{\apj}
  {\bf 412} (July, 1993) 64--71}.

\bibitem{Ross12}
A.~J. {Ross}, W.~J. {Percival}, A.~G. {S{\'a}nchez}, L.~{Samushia}, S.~{Ho},
  E.~{Kazin} et~al., \emph{{The clustering of galaxies in the SDSS-III Baryon
  Oscillation Spectroscopic Survey: analysis of potential systematics}},
  \href{http://dx.doi.org/10.1111/j.1365-2966.2012.21235.x}{\emph{\mnras} {\bf
  424} (July, 2012) 564--590}, [\href{http://arxiv.org/abs/1203.6499}{{\tt
  1203.6499}}].

\bibitem{Limber53}
D.~N. {Limber}, \emph{{The Analysis of Counts of the Extragalactic Nebulae in
  Terms of a Fluctuating Density Field.}},
  \href{http://dx.doi.org/10.1086/145672}{\emph{\apj} {\bf 117} (Jan., 1953)
  134}.

\bibitem{Eisenstein07}
D.~J. {Eisenstein}, H.-J. {Seo}, E.~{Sirko} and D.~N. {Spergel},
  \emph{{Improving Cosmological Distance Measurements by Reconstruction of the
  Baryon Acoustic Peak}}, \href{http://dx.doi.org/10.1086/518712}{\emph{\apj}
  {\bf 664} (Aug., 2007) 675--679},
  [\href{http://arxiv.org/abs/astro-ph/0604362}{{\tt astro-ph/0604362}}].

\bibitem{Padmanabhan12}
N.~{Padmanabhan}, X.~{Xu}, D.~J. {Eisenstein}, R.~{Scalzo}, A.~J. {Cuesta},
  K.~T. {Mehta} et~al., \emph{{A 2 per cent distance to z = 0.35 by
  reconstructing baryon acoustic oscillations - I. Methods and application to
  the Sloan Digital Sky Survey}},
  \href{http://dx.doi.org/10.1111/j.1365-2966.2012.21888.x}{\emph{\mnras} {\bf
  427} (Dec., 2012) 2132--2145}, [\href{http://arxiv.org/abs/1202.0090}{{\tt
  1202.0090}}].

\bibitem{Burden14}
A.~{Burden}, W.~J. {Percival}, M.~{Manera}, A.~J. {Cuesta}, M.~{Vargas Magana}
  and S.~{Ho}, \emph{{Efficient reconstruction of linear baryon acoustic
  oscillations in galaxy surveys}},
  \href{http://dx.doi.org/10.1093/mnras/stu1965}{\emph{\mnras} {\bf 445} (Dec.,
  2014) 3152--3168}, [\href{http://arxiv.org/abs/1408.1348}{{\tt 1408.1348}}].

\end{thebibliography}\endgroup


\end{document}